\documentclass[a4paper,11pt]{article}
\usepackage{packages}
\usepackage{pos}

\title{Performance of two-level sampling for the glueball spectrum in pure gauge theory}

\author*[a]{Lorenzo Barca}
\author[b]{Francesco Knechtli}
\author[c]{Michael Peardon}
\author[a]{Stefan Schaefer}
\author[b]{Juan Andr\'es Urrea-Ni\~{n}o}

\affiliation[a]{John von Neumann-Institut f\"ur Computing NIC, Deutsches Elektronen-Synchrotron DESY, Germany}
\affiliation[b]{Department of Physics, University of Wuppertal, Gaußstrasse 20, 42119 Germany}
\affiliation[c]{School of Mathematics, Trinity College Dublin, Ireland}

\emailAdd{lorenzo.barca@desy.de}

\abstract{
The computation of the glueball spectrum is particularly challenging due to the rapid decay of the signal-to-noise ratio of the correlation functions. To address this issue, advanced techniques such as gauge link smearing and the variational method are commonly employed to identify the spectrum before the signal diminishes significantly. However, a significant improvement in the signal-to-noise ratio can be achieved by utilising multilevel sampling techniques. In this talk, we present a study of the glueball spectrum in pure gauge theory with a two-level algorithm. Specifically, we explore the relation between noise reduction and the various multilevel parameters, such as the width of the dynamical regions and the number of submeasurements.}

\FullConference{The 40th International Symposium on Lattice Field Theory (Lattice 2023)\\
July 31st - August 4th, 2023\\
Fermi National Accelerator Laboratory\\}

\begin{document}
\maketitle
\section{Introduction}
One of the major challenges of lattice QCD simulations is the signal-to-noise ratio (S/N) of the correlation functions. 
Particularly demanding, even in pure gauge theory, is the computation of the spectrum of glueballs, hypothetical 
composite particles made of solely gluons. The effective masses are extracted from the glueball correlation functions, whose signal-to-noise ratio decays exponentially over time with the glueball mass \cite{Parisi:1983ae, Lepage:1989hd}. 
Since the S/N prevents one from reaching the large time distances, where eventually the ground state dominates, a variational method is adopted to extract the effective masses from the solutions of a generalised eigenvalue problem (GEVP) \cite{MICHAEL1983433}. 
The variational basis is constructed with Wilson loops of different shapes, see \cite{Berg:1982kp}, and in addition, fuzzing 
techniques such as link smearings are applied to mitigate UV fluctuations and construct glueball operators that overlap 
better with the ground state. Adopting these standard techniques, different lattice groups have managed to compute the 
glueball spectrum in pure gauge theory with good consistency \cite{Morningstar:1999rf, Bali:1993fb, Athenodorou:2020ani}. 
In particular, in \cite{Athenodorou:2020ani}, the statistical error was reduced by a factor of $10$ with respect to previous 
works by considering a large basis of operators and very high-statistics. 

In this work, we directly address the S/N problem by using a more efficient algorithm for the computation of the glueball 
correlation functions and we adopt a variational method to extract the effective masses. In lattice gauge theory, 
a relevant error reduction was first achieved in \cite{PARISI1983418} for the computation of the string tension from 
the correlation functions of Polyakov loops, and it was known as the `multihit' method. In \cite{Luscher:2001up}, 
the algorithm was improved by exploiting locality of the pure gauge theory and the factorisation properties of the 
traces of Polyakov loops, which yelded an exponential error reduction. This algorithm was named multilevel. 
In \cite{Meyer:2002cd, Meyer:2003hy}, in the same spirit as in \cite{PARISI1983418, Luscher:2001up}, a two-level algorithm 
was succesfully applied for the computation of glueball masses in 3+1 SU(3) and 2+1 SU(2) Yang-Mills (YM) theory, respectively.

In our study, we use an algorithm similar to the one used in \cite{Meyer:2002cd, Meyer:2003hy}, to investigate SU(3) pure gauge theory 
in 3+1 dimensions and in particular, to study the performance of the multilevel sampling algorithm with respect to its parameters.

\section{Multilevel algorithm}
\subsection{Comparison between traditional and multilevel algorithm}
The glueball correlation functions are defined as
\begin{equation}\label{gb_corr_std}
	C( t, t_0) = \langle W^{\Gamma}( t + t_0) W^{\Gamma}(t_0) \rangle ~
\end{equation}
where $W^{\Gamma}(t)$ is the glueball operator and $\langle \cdots \rangle$ represents the gauge average.
The glueball operator $W^{\Gamma}(t)$ is constructed from a space-like Wilson loop projected on a specific channel ${\Gamma} = R^{PC}$, which contains the lattice quantum numbers\footnote{In the rest frame, the symmetry group is the octahedral group $O_h$ and the lattice irreps are $A_1$, $A_2$, $E$, $T_1$ and $T_2$.} (irreducible representation $R$, parity $P$, and charge conjugation $C$). 
The correlation functions in eq.~\eqref{gb_corr_std} is estimated with Monte Carlo techniques by measuring the traces of Wilson loops on an ensemble of $N_{\mathrm{cfg}}$ gauge configurations $U(\vec{x}, t)$, so that
\begin{equation}\label{gb_corr_hmc}
	C(t, t_0) = \frac{1}{N_{\mathrm{cfg}}}
	\sum_{i} \mathrm{Tr}\left[ W^{\Gamma}(U_i,  t + t_0) W^{\Gamma}(U_i,  t_0) \right]
	+ \mathcal{O}\left(1 / \sqrt{N_{\mathrm{cfg}}} \right)~.
\end{equation}

The error thus scales as $\propto 1 / \sqrt{N_{\mathrm{cfg}}}$ with the standard method.
Instead, with a multilevel algorithm the glueball correlation functions are evaluated with
\begin{equation}\label{gb_corr_mlvl}
	C( t, t_0) = \langle \left[ W^{\Gamma}( t + t_0)\right]_{\lambda_1} \left[ W^{\Gamma}(t_0)\right]_{\lambda_2} \rangle ~,
\end{equation}
where the glueball operators $W^{\Gamma}(t)$ are submeasured additionally onto two different subsets $\lambda_1$ and $\lambda_2$ of the full temporal lattice extent with lattice sites $t/a=0, \dots, N_t -1$. 
These subsets $\lambda_k$ are the dynamical regions of the temporal lattice, which are bounded by frozen regions $\phi_{jk}$. In other words, we generate $N_{\mathrm{cfg}}$ gauge configurations $U(\vec{x}, t)$ which are well thermalised, and then each gauge configuration is additionally updated only on the timeslices that belong to specific subsets $\lambda_k$ of the full temporal lattice, while they are kept fixed on the timeslices that belong to the frozen regions $\phi_{jk}$. Examples of such subsets are displayed in Fig.~\ref{Figure:mlvl_sublattices}.
The parameters that define the multilevel algorithm are the number of submeasurements $N_2$ performed on the different dynamical regions $\lambda_k$ and the width of each of these regions.
The advantage of using the multilevel algorithm is that the statistical error is reduced by a factor of $N_2$ by doing $N_2$ more submeasurements (labelled by $\left[\cdots\right]$) with respect to the gauge average only (labelled by $\langle \cdots \rangle$), provided that the gauge configurations are updated independently in the two regions $\lambda_j$ and $\lambda_k$ and that the distance from the boundaries $\phi_{jk}$ is large enough, as we will discuss in the next section.
The Wilson loop is projected onto zero momentum by summing over all the spatial lattice sites.
The brackets in eq.~\eqref{gb_corr_std} refer to the gauge average of the product of traces of Wilson loops.

\subsection{This work: ensemble generation}
We consider a pure $\mathrm{SU}(3)$ YM theory in $3+1$ dimensions. The lattice discretised action can be written in terms of the plaquettes and it reads
\begin{equation}
S_g  = \frac{\beta}{3}
\sum_{n \in \Lambda}
\sum_{\mu < \nu}
\mathrm{Re}
\left\{
\mathrm{Tr}
\left[ 1 - U_\mu(n) 
U_\nu(n+\hat{\mu})
U_\mu(n+\hat{\nu})^{\dagger} U_\nu(n)^{\dagger} 
\right]
\right\}~.
\end{equation}

We investigate a single ensemble with $V/a^4=24\times 24^3$ and with periodic boundary conditions at $\beta=6.2$, whose lattice spacing corresponds to $a=0.0677~\mathrm{fm}$, according to \cite{Necco:2001xg} with $r_0 = 0.5~\mathrm{fm}$. In \cite{Sakai:2022zdc}, results for the glueball spectrum on a similar ensemble were obtained with the spatial gradient flow, which we refer to for a comparison. At first, we generate $\mathcal{O}(10^4)$ gauge configurations with the traditional HMC algorithm with a trajectory step of $\tau=3$ and we study the autocorrelation time of the plaquette and the topological charge at different flow times.
We observe that with a spacing of $100\tau$ the autocorrelations of the gauge configurations 
are negligible, i.e., the integrated autocorrelation time for the topological charge is $\tau_{\mathrm{int}} \approx 0.5$ \cite{Wolff:2003sm}.

\subsection{Multilevel updates}

\begin{figure}[t!]
	\includegraphics[width=\textwidth]{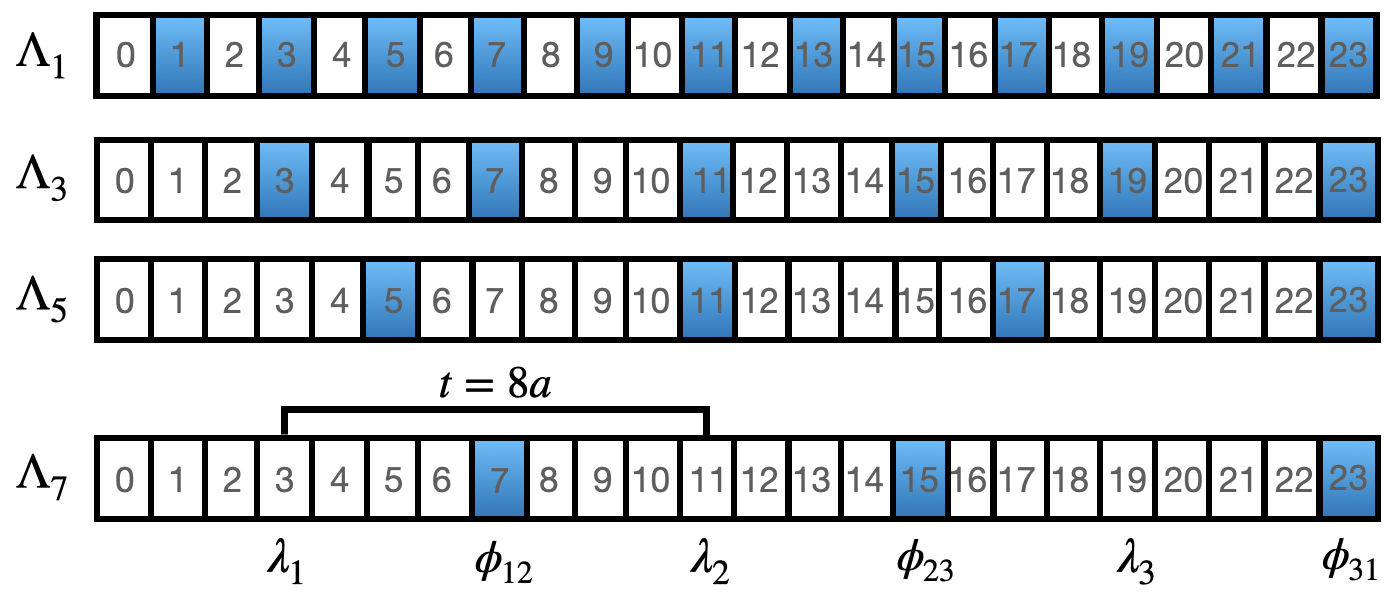}
	\caption{Some of the different decompositions that are investigated in this preliminary work, where the temporal lattice extent is $N_t=24$. 
		The blue cells represent the fixed boundaries on the second level, whereas the white cells are the dynamical ones. For $\Lambda_1$, the gauge configurations are additionally updated on the even timeslices and kept fixed on the odd timeslices. This yelds $12$ dynamical regions $\lambda$ and $12$ frozen regions $\phi$. On the bottom, an example of a correlation between Wilson loops submeasured on two different dynamical regions $\lambda_1$ and $\lambda_2$ of $\Lambda_7$, i.e., $\langle \left[W(t_0 +  t =11a)\right]_{\lambda_1} \left[W(t_0=3a)\right]_{\lambda_2} \rangle$.}
	\label{Figure:mlvl_sublattices}
\end{figure}

\begin{figure}[t!]
	\includegraphics[width=\textwidth]{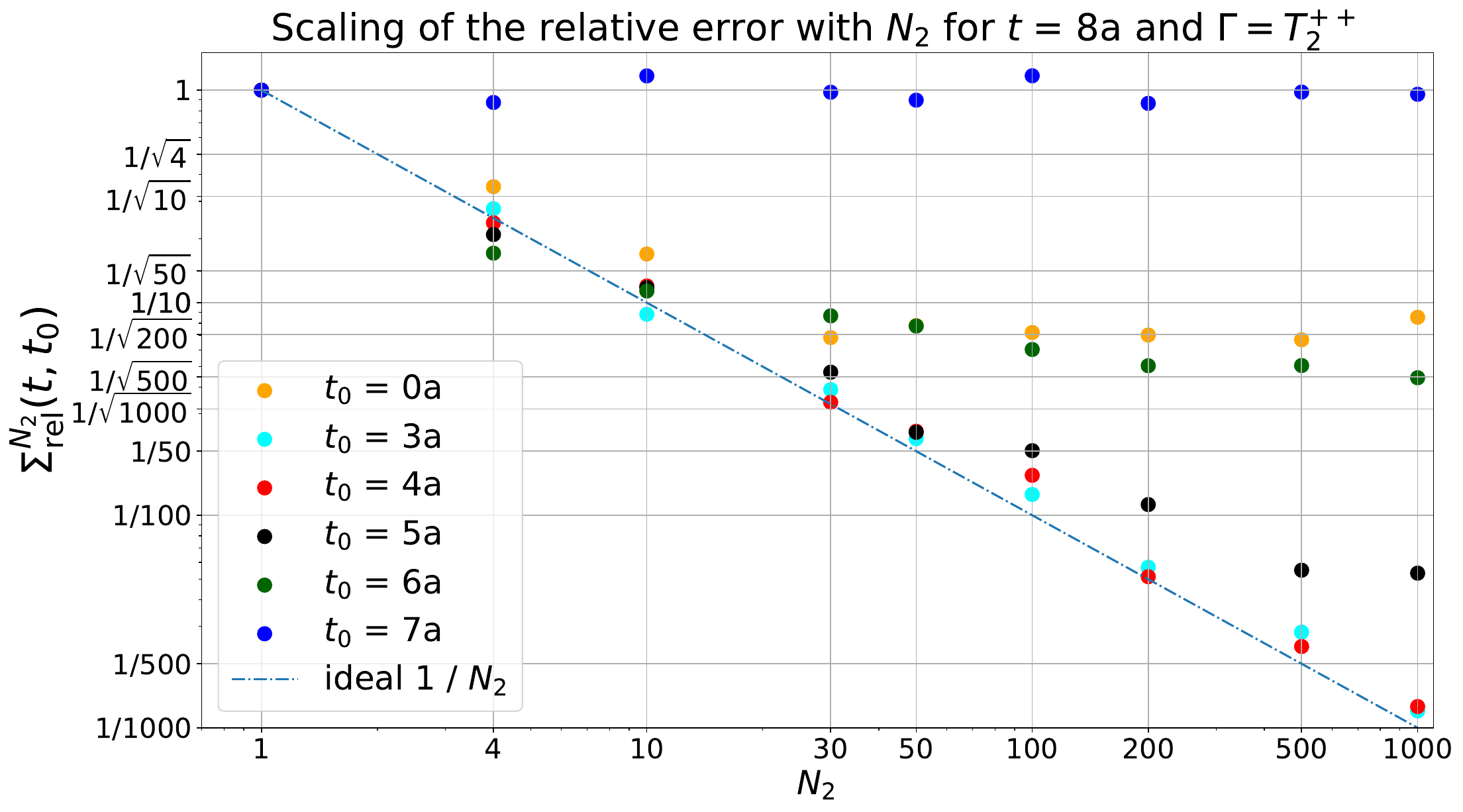}
	\caption{Comparison plot of $\Sigma^{N_2}_{\mathrm{rel}}(t,t_0) = \sigma^{N_2}(t, t_0)/\sigma^{N_2=1}(t, t_0)$ for the channel $\Gamma=T_2^{++}$ at fixed $t=8a$, 
	and at different $t_0$, to show that the multilevel works best with Wilson loops that are quite far from the boundaries. Only correlations between operators at central timeslices like $t_0=3a$ to $t=11a$ (visualise in Fig.~\ref{Figure:mlvl_sublattices}) achieve the expected error reduction, close to the ideal scaling $1/N_2$. The error reduction decreases as the Wilson loops get closer to the boundaries ($\phi = \{7a, 15a, 23a\}$ for this lattice decomposition $\Lambda_7$). There is an error on the error which is not displayed here and might be relevant at large $N_2$, see \cite{Wolff:2003sm}.}
	\label{Figure:er_teq8}
\end{figure}

In order to make more submeasurements on local regions $\left[\dots \right]_{\lambda_k}$, we perform further updates of the gauge configurations by considering dynamical regions of increasing size as depicted in Fig.~\ref{Figure:mlvl_sublattices}, 
and we compute a glueball operator $W^{\Gamma}(t) ~ \forall ~t \in \lambda_k, \phi_k$. 
The different decompositions of the temporal lattice are labelled as $\Lambda_w$, where $w$ represents the width in lattice units of the dynamical regions $\lambda_k$. 
Clearly, with the decomposition $\Lambda_1$, where the dynamical regions are only one lattice spacing wide, one would expect a reduction of the statistical error just for the even timeslices, as the gauge configurations are not updated on the odd timeslices. 
The reduction of the statistical error $\sigma^{N_2}(t,t_0)$ with $N_2$ submeasurements depends on the distance from the boundaries (blue cells in Fig.~\ref{Figure:mlvl_sublattices}), 
so that, if we consider for example the lattice decomposition $\Lambda_7$, $\sigma^{N_2}(t=8a, t_0=3a) < \sigma^{N_2}(t=8a, t_0=6a)$ as $t_0 = 3a$ and $t+t_0=11a$ are more distant from the boundaries with respect to correlations between $t_0=6a$ and $t_0 + t=14a$.
In Fig.~\ref{Figure:er_teq8}, we present the results of the error reduction $\Sigma^{N_2}_{\mathrm{rel}}(t, t_0)$ at $t = 8a$ for the glueball correlation functions projected on the channel $T_2^{++}$ and with the lattice decomposition $\Lambda_7$. 
The error reduction is defined as the ratio of the statistical error with $N_2$ and with $N_2=1$ submeasurements, so to show the expected scaling $1/N_2$, and it is plotted against the number of submeasurements $N_2$.
As can be seen from the plot in Fig.~\ref{Figure:er_teq8}, the expected error reduction of $1/N_2$ is almost perfect for $t_0 = 3a, ~4a$ and $t=8a$, 
whereas the error reduction for $t_0=6a$ and $t_0=0a$ starts to saturate at around $N_2=30$.
For correlations between Wilson loops evaluated at the boundaries like $\langle W(t_0 + t=15a) W(t_0 = 7a) \rangle$, 
no error reduction is observed as expected from the fact that the gauge configurations are not updated on these timeslices.
The larger the dynamical regions, the better multilevel performance, thus explaining why a larger dynamical region is a better choice.
Similar behaviour is observed for correlations of Wilson loops at different $t_0$ and $t$ for the channels $T_2^{++}$, and similarly for $E^{++}$.
A particular channel is the $A_1^{++}$, where the $A_1^{++}$ Wilson loop has nonvanishing vacuum expectation value (VEV), i.e. $\langle W^{A_1^{++}}(t) \rangle \neq 0  ~\forall ~ t$, which must be subtracted carefully as discussed in \cite{Meyer:2003hy}. 
For this channel, we do not observe an error reduction as large as for other channels $E^{++}$, and $T_2^{++}$. This was observed also in \cite{Meyer:2003hy}.
However, the signal-to-noise ratio is not as bad as for other channels, provided that few smearing iterations are applied.

\subsection{Multilevel analysis}
With the multilevel algorithm, the correlation functions $C(t, t_0)$ have different variances for different $t$ and $t_0$, as one can see, for instance, from the relative statistical error plotted in Fig.~\ref{Figure:er_teq8}. 
For this reason, the sum over all the equivalent $t_0$ can be replaced by a weighted average
\begin{equation}\label{weighted_avrg}
\widebar{C}(t) = 
\frac{\sum_{t_0} w(t, t_0) C(t_0 + t, t_0)}
{\sum_{t_0 } w(t, t_0)}~.	
\end{equation}

In \cite{Meyer:2003hy}, the weights $w(t, t_0)$ were chosen to minimise the final statistical error. In this work, we consider the inverse of the variance, i.e., $1/\sigma^2(t, t_0)$, but one can consider a more sophisticated weighting procedure to minimise the final statistical error.

\section{Variational method}

\begin{figure}[t!]
	\includegraphics[width=\textwidth]{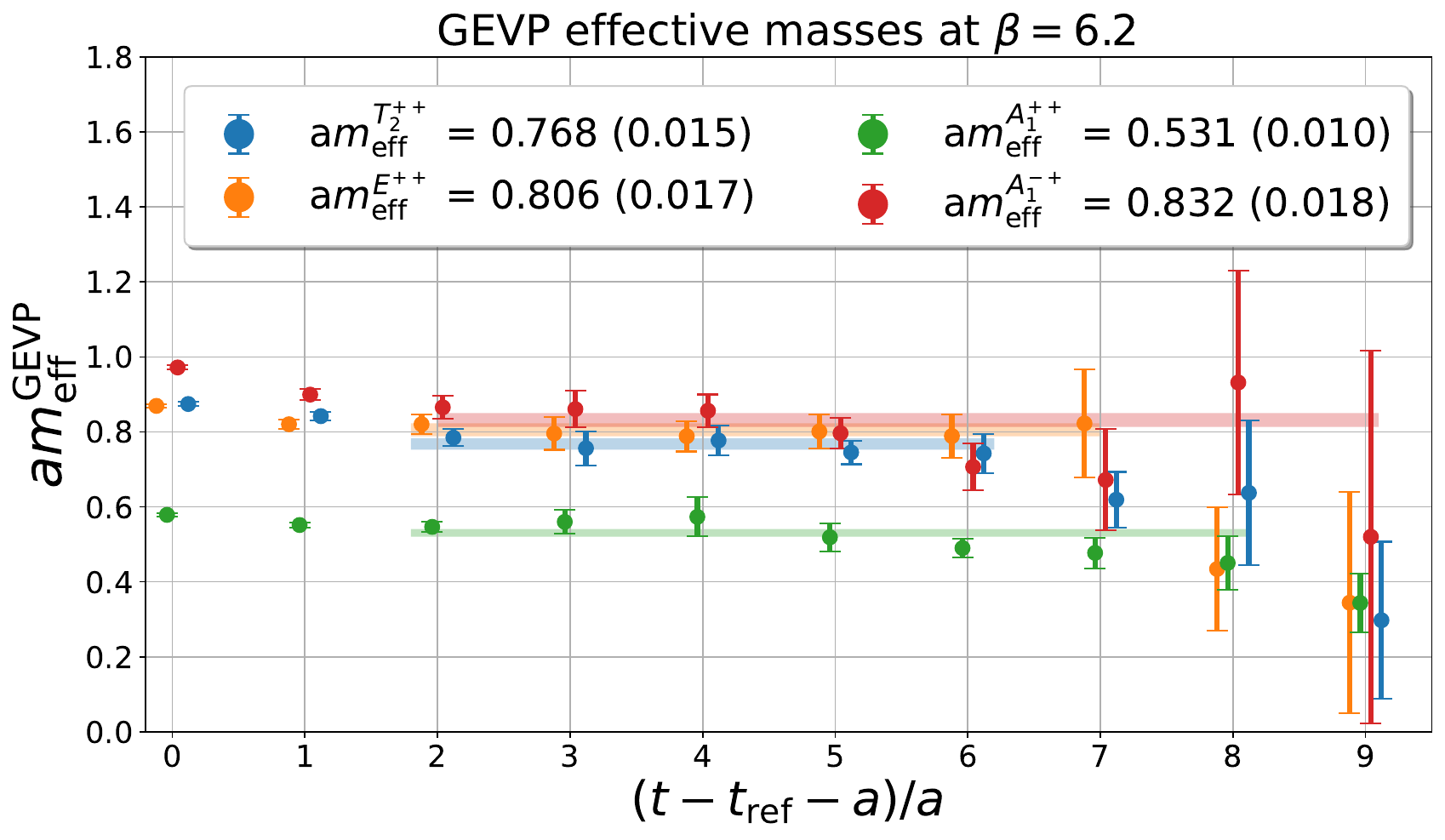}
	\caption{GEVP effective masses at $t_{\mathrm{ref}}=0a$ for the channels $A_1^{++}$, $E^{++}$, $T_2^{++}$, and $A_1^{-+}$ at $\beta=6.2$, with $V/a^4=24\times 24^3$. These results are obtained with the sublattice decomposition $\Lambda_{11}$, where the full temporal lattice with $N_t=24a$ is split into two dynamical regions of width $w=11a$.}
	\label{Figure:gevp_results_meff}
\end{figure}

We construct a basis of $5$ different length-$8$ Wilson loops, which can be projected onto the channels of interest $A_1^{++}$, $E^{++}$, $T_2^{++}$, and $A_1^{-+}$. 
Each of these Wilson loops is smeared with $4$ different APE smearing levels to construct an efficient variational basis. For each $W^{\Gamma}_j(t)$ in this basis, we compute
\begin{equation}
C^{\Gamma}_{ij}(t, t_0) = 
\langle W^{\Gamma}_i( t + t_0) W^{\Gamma}_j(t_0) \rangle ~,
\hspace{6em}
i,j = 1,\dots, N
\end{equation}
and we estimate the weighted average $\widebar{C}^{\Gamma}_{ij}(t)$ as discussed in the previous section, see eq.~\eqref{weighted_avrg}. 
We then solve the GEVP
\begin{equation}
\widebar{C}^{\Gamma}(t) V^{\Gamma}(t, t_{\mathrm{ref}}) = 
\widebar{C}^{\Gamma}(t_{\mathrm{ref}}) \Lambda^{\Gamma}(t, t_{\mathrm{ref}}) V^{\Gamma}(t, t_{\mathrm{ref}})
~
\hspace{3em}
\forall ~t > t_{\mathrm{ref}}~,
\end{equation}
where $\Lambda^{\Gamma}(t,t_{\mathrm{ref}})=\mathrm{diag}\left(\lambda^{\Gamma}_1(t, t_{\mathrm{ref}}), \dots, \lambda^{\Gamma}_N(t, t_{\mathrm{ref}}\right)$ and $V^{\Gamma}(t,t_{\mathrm{ref}})=\left(v^{\Gamma}_1(t, t_{\mathrm{ref}}), \dots, v^{\Gamma}_N(t, t_{\mathrm{ref}})\right)$ are the matrices of generalised eigenvalues and eigenvectors, respectively.
The GEVP effective masses $m^{\Gamma}_{\mathrm{eff}}$ can be extracted from the eigenvalues as $\lambda^{\Gamma}_k(t, t_{\mathrm{ref}}) \propto e^{-m^{\Gamma}_{\mathrm{eff}}{(t- t_{\mathrm{ref}})}}$ for the channels $\Gamma=A_1^{++}, E^{++}, T_2^{++}, A_1^{-+}$.
Some operators with different shapes might be degenerate at large enough smearing radii, as observed in Fig.~8 of \cite{Sakai:2022zdc}, and thus some careful choice must be made for the optimal variational basis. A possibility to construct the optimal variational basis is to use the singular value decomposition of the GEVP matrix $\widebar{C}(t)$ and prune the matrix, or solve the GEVP with different combinations of the operators in the variational basis and observe empirically which basis works best.
The choice of the variational basis is very important even with very high statistics, and in this preliminary study we focus only on large Wilson loops as they contain all the irreps of interest. Since large loops comprise of more gauge links, statistical fluctuations are expected to be larger with respect to, for example, a $1\times1$ Wilson loop (plaquette), which is of length-$4$, but it might also be that larger Wilson loops have greater overlap onto the ground state.
We adopt the two-level algorithm on $N_{\mathrm{cfg}}=101$ configurations with up to $N_2=1000$ submeasurements to compute each of these correlation functions and solve the GEVP. 
The two-level decomposition that works best is the one with larger dynamical region, and we decompose the full temporal lattice with $t=0a, ~\dots, ~23a$ into two dynamical regions of width $11a$, separated by a frozen region or fixed boundary\textbf{} that consists of a single timeslice.
The effective masses of the ground state for the channels $\Gamma = A_1^{++}$, $E^{++}$, $T_2^{++}$, and $A_1^{-+}$, are extracted from the exponential decay of the eigenvalues and are plotted in Fig.~\ref{Figure:gevp_results_meff}. The results are consistent within the error with \cite{Sakai:2022zdc} and will improve with a better choice of operators and smearing radii.

\section{Conclusions}
In this preliminary work, we analyse the $SU(3)$ glueball spectrum in $3+1$ dimensions with a two-level algorithm at $\beta=6.2$ and $V/a^4=24\times24^3$. 
We consider $101$ gauge configurations and perform up to $N_2=1000$ submeasurements on the inner level. We achieve an error reduction proportional to $N_2$ for some correlations that are at least $3$ lattice spacings away from the boundaries for the channels $E^{++}$, $T_2^{++}$, and $A_1^{-+}$.
For the irrep $A_1^{++}$, which contains the lighest glueball and a nonvanishing vacuum expectation value, the error reduction is smaller than $N_2$, but the signal-to-noise ratio is less challenging. 
The results are consistent within the errors with literature and are very promising in terms of the error reduction. 
In the future, we will address the question of the scaling of the multilevel towards the continuum limit and the different channels. In addition, we plan to show a direct comparison with the traditional method to highlight the benefit in computer time of using the multilevel sampling algorithm compared to the traditional method.

\paragraph{Acknowledgements}
The work is supported by the German Research Foundation (DFG) research unit FOR5269 "Future methods for studying confined gluons in QCD". The authors thank all the members of the research unit for useful discussions. The authors gratefully acknowledge the
scientific support and HPC resources provided by the Erlangen National High Performance Computing Center (NHR@FAU) of the Friedrich-Alexander-Universität Erlangen-Nürnberg (FAU) under the NHR project k103bf.

\bibliography{bibliography/bibliography}

\begin{thebibliography}{10}

\bibitem{Parisi:1983ae}
G.~Parisi.
\newblock {The Strategy for Computing the Hadronic Mass Spectrum}.
\newblock {\em Phys. Rept.}, 103:203--211, 1984.

\bibitem{Lepage:1989hd}
G.~Peter Lepage.
\newblock {The Analysis of Algorithms for Lattice Field Theory}.
\newblock In {\em {Theoretical Advanced Study Institute in Elementary Particle
  Physics}}, 6 1989.

\bibitem{MICHAEL1983433}
C.~Michael and I.~Teasdale.
\newblock Extracting glueball masses from lattice qcd.
\newblock {\em Nuclear Physics B}, 215(3):433--446, 1983.

\bibitem{Berg:1982kp}
B.~Berg and A.~Billoire.
\newblock {Glueball Spectroscopy in Four-Dimensional SU(3) Lattice Gauge
  Theory. 1.}
\newblock {\em Nucl. Phys. B}, 221:109--140, 1983.

\bibitem{Morningstar:1999rf}
Colin~J. Morningstar and Mike~J. Peardon.
\newblock {The Glueball spectrum from an anisotropic lattice study}.
\newblock {\em Phys. Rev. D}, 60:034509, 1999.

\bibitem{Bali:1993fb}
G.~S. Bali, K.~Schilling, A.~Hulsebos, A.~C. Irving, Christopher Michael, and
  P.~W. Stephenson.
\newblock {A Comprehensive lattice study of SU(3) glueballs}.
\newblock {\em Phys. Lett. B}, 309:378--384, 1993.

\bibitem{Athenodorou:2020ani}
Andreas Athenodorou and Michael Teper.
\newblock {The glueball spectrum of SU(3) gauge theory in 3 + 1 dimensions}.
\newblock {\em JHEP}, 11:172, 2020.

\bibitem{PARISI1983418}
G.~Parisi, R.~Petronzio, and F.~Rapuano.
\newblock A measurement of the string tension near the continuum limit.
\newblock {\em Physics Letters B}, 128(6):418--420, 1983.

\bibitem{Luscher:2001up}
Martin Luscher and Peter Weisz.
\newblock {Locality and exponential error reduction in numerical lattice gauge
  theory}.
\newblock {\em JHEP}, 09:010, 2001.

\bibitem{Meyer:2002cd}
Harvey~B. Meyer.
\newblock {Locality and statistical error reduction on correlation functions}.
\newblock {\em JHEP}, 01:048, 2003.

\bibitem{Meyer:2003hy}
Harvey~B. Meyer.
\newblock {The Yang-Mills spectrum from a two level algorithm}.
\newblock {\em JHEP}, 01:030, 2004.

\bibitem{Necco:2001xg}
Silvia Necco and Rainer Sommer.
\newblock {The N(f) = 0 heavy quark potential from short to intermediate
  distances}.
\newblock {\em Nucl. Phys. B}, 622:328--346, 2002.

\bibitem{Sakai:2022zdc}
Keita Sakai and Shoichi Sasaki.
\newblock {Glueball spectroscopy in lattice QCD using gradient flow}.
\newblock {\em Phys. Rev. D}, 107(3):034510, 2023.

\bibitem{Wolff:2003sm}
Ulli Wolff.
\newblock {Monte Carlo errors with less errors}.
\newblock {\em Comput. Phys. Commun.}, 156:143--153, 2004.
\newblock [Erratum: Comput.Phys.Commun. 176, 383 (2007)].

\end{thebibliography}
\bibliographystyle{unsrt}

\end{document}